\documentclass[aps,preprint,superscriptaddress,nofootinbib,a4paper]{revtex4}
\usepackage[utf8]{inputenc} 
\usepackage[T1]{fontenc} 
\usepackage[dvipsnames]{xcolor}
\usepackage{epsf, epsfig, graphicx,wrapfig,subfigure,float,url,ulem,enumitem,footnotebackref}
\usepackage{amsmath, amssymb, amsthm, physics, braket,mathrsfs,nccbbb,cancel}
\usepackage[toc,page]{appendix}
\usepackage[hang, flushmargin]{footmisc}
\usepackage[mathscr]{eucal}
\usepackage{color}

\newcommand{\be}{\begin{equation}}
\newcommand{\ee}{\end{equation}}

\begin{document}

\title{The Influence of Angular Momentum and Chemical Potential on Holographic Entanglement Entropy}

\author{Po-Chun Sun}\email{jim.pochun.sun@gmail.com}
\affiliation{Department of Physics, National Central University, Chungli 32001, Taiwan.}
\date{July 16, 2021}

\begin{abstract}
We study the entanglement entropy between a strip region with width $2R$ and its complement in strongly coupled large-$N$ conformal field theory (CFT) on $\mathbb{R}^{1,n}$ with chemical potential and angular momentum in an thermal equilibrium state, which dual to the cylindrical Kerr-Newman black hole. Using Ryu-Takanayagi conjecture, we are able to explore the entanglement entropy through calculating the area of the extremal surface which is anchored on the entangled surface. Because we consider the rotating charged black hole for the gravitational dual, the entropy can be characterized by the mass $m$, charge $q$ and rotation $a$ parameters. We find that in the small $R$ limit, only the mass $m$ and the rotation $a$ parameters come in the leading behavior, the quadratic of $R$ term, after subtracting the entanglement entropy in vacuum. In the large $R$ limit, the area of the extremal surface is very close to the area of the event horizon after regularization. When we take the whole spatial region, the entanglement entropy is identical to the thermal entropy, i.e. the Bekenstein-Hawking entropy.
\end{abstract}


\maketitle

\tableofcontents

\section{Introduction}
Entanglement is a quantum measurement with non-local property, which plays an important role in quantum information. Let's consider a quantum system which can be divided into two parts, $A$ and $B$, such that the total Hilbert space is a direct product of these two subsystems as $H_\mathrm{tot} = H_{A} \otimes H_{B}$. For a quantum state $\ket{\Psi} \in H_\mathrm{tot}$ the reduced density matrix of $A$ is defined, by tracing out the degree of freedom of subsystem $B$, as $\rho_A = \tr_B (\ket{\Psi} \bra{\Psi})$. Then, the entanglement entropy(von Neumann entropy) of $A$ is defined as
\begin{equation} 
\mathcal{S}_{A} \equiv - \tr(\rho_{A} \, \ln \rho_{A}),
\end{equation}
which is the measurement to describe the correlation between $A$ and $B$.

On the other hand, in many examples, the holographic principle~\cite{tHooft:1993dmi, Susskind:1994vu} has successfully connected two very different theories: gravity in the bulk and quantum field theory (QFT) on the boundary. One of the famous examples is the AdS/CFT correspondence~\cite{Maldacena:1997re, Gubser:1998bc, Witten:1998qj}. In this framework, Ryu and Takanayagi proposed that the entanglement entropy in QFT between the spatial region $\Sigma$ and its outside can be computed from~\cite{Ryu:2006bv, Ryu:2006ef}
\begin{equation} 
\mathcal{S}_{\Sigma} = \frac{\mathcal{A}_{\Gamma}}{4},
\end{equation}
where $\mathcal{A}_{\Gamma}$ is the area of the extremal surface (with minimum area) in the bulk that the end of which is connected to the entangled surface.

The Ryu-Takayanagi proposal provides a very different and, in many situations, convenient approach to study the entanglement entropy. The calculations of the holographic entanglement entropy in vacuum were firstly carried out in~\cite{Ryu:2006bv, Ryu:2006ef}. The references~\cite{Nishioka:2009un, Takayanagi:2012kg, Rangamani:2016dms} provide nice reviews about studying entanglement entropy by the holographic method. Although the entanglement entropy is very different from the thermal entropy, people have been analyzed the thermodynamics of entanglement entropy in the thermal state by exploring the holographic entanglement entropy of black holes in the small $R$ limit~\cite{Bhattacharya:2012mi,Fischler:2013, Chaturvedi:2016kbk, Karar:2018ecr, Saha:2019ado, Caputa:2013lfa,Kundu:2016,Nadi:2019bqu,Maulik:2021,Singh:2021}. Moreover, it has been proved that the extremal surface cannot penetrate the horizon in static black holes~\cite{Hubeny:2012ry} which implies that in the large $R$ limit, the area of extremal surface will be dominated by the IR region after regularization due to a great part of the extremal surface runs along the event horizon. One can expect that, after regularization, the leading behavior of the area of the extremal surface in the large $R$ limit is identical to Bekenstein-Hawking entropy, the thermal entropy. There are some discussions of the entanglement entropy in the large $R$ limit~\cite{Fischler:2013,Caputa:2013lfa,Liu:2013una, Chaturvedi:2016kbk,Kundu:2016,Saha:2019ado, Karar:2018ecr}.

The entanglement entropy is always divergent in quantum field theories if we take the continuous limit. The leading behavior of entanglement entropy is proportional to the volume of the entangling surface owing to short-range correlations, it is so-called area law~\cite{Bombelli:1986,Srednicki:1993,Eisert:2010}, and the terms which are independent of cutoff are attributed to large-range entanglement. These finite terms of the entanglement entropy in a CFT are related to the central charge, featuring the number of degrees of freedom. On top of that, the holographic entanglement entropy is the fine-grained entropy, the definition of which is full quantum at a microscopic level, on the other hand, Bekenstein-Hawking entropy, the thermal entropy, is the coarse-grained entropy, which bases on effectively semi-classical. Thus, We expect the holographic entanglement entropy of black hole is very different from the thermal entropy for the small $R$ limit. As we take the large $R$ limit, the two entropies are approximately the same. 

In this paper, we work on the holographic entanglement entropy between a strip region with width $2R$ and its complement in $(n + 1)$-dimensional strongly coupled large-$N$ CFT on $\mathbb{R}^{1,n}$ with chemical potential and angular momentum in the small and large $R$ limits. The gravitational dual is an $(n + 2)$-dimensional cylindrical Kerr-Newman black hole, which the boundary of the spatial region is topological flat $\mathbb{R}^n$. In section \ref{sec2}, we introduce the cylindrical Kerr-Newman-AdS black hole metric, and highlight several important properties in such spacetime. Moreover, we build up the formula of the area when $\Sigma$ is a strip in section \ref{sec3}. Although there is an rotation in $\phi$ direction, the extremal surface $\Gamma$ still preserved the translational symmetry along $\phi$ direction, and the corresponding constant of motion can simplify our calculations. However, it is still difficult to get the analytic results in general case. Here, we only focus on the small and large $R$ limits, the derivation of which in section \ref{sec4} and \ref{sec5} respectively. In the end, we summarize our analytical results and briefly discuss the outlook in \ref{sec6}.

\section{Cylindrical Kerr-Newman-AdS Black Hole}\label{sec2}
Let's consider an observer moves with the velocity $\Omega= a/\Theta$ in an $(n + 2)$-dimensional charged AdS-black branes, that is, Lorentz boost along the spatial direction $\phi$
\begin{equation} \label{ct}
dt \; \rightarrow \; \Theta \, dt - a \, d\phi, \qquad d\phi \; \rightarrow \; \Theta \, d\phi - a \, dt,
\end{equation}
where $a \in (-\infty, \infty)$ is a rotation parameter when the spatial coordinate $\phi$ is identified as $\phi \sim \phi + 2\pi$ and $\Theta = \sqrt{1 + a^2}$. The line element of a cylindrical Kerr-Newman black brane becomes~\cite{Awad:2002cz}
\begin{equation} \label{metric}
ds^{2} = \frac{L^2}{z^2} \left[ - h(z) (\Theta \, dt - a \, d\phi)^2 + \frac{dz^2}{h(z)} + (\Theta \, d\phi - a \, dt)^2 + d\sigma_{n-1}^2 \right],
\end{equation}
where $d\sigma_{n-1}^2$ is the line element of $(n-1)$-dimensional Euclidean space and the emblackening function is
\begin{equation} \label{hmq}
h(z) = 1 - m z^{n+1} + q^2 z^{2 n},
\end{equation}
including mass and charge parameters. The associated gauge potential is
\begin{equation}
A = \sqrt{\frac{n}{2 (n - 1)}} \, q \, (z_h^{n-1} - z^{n-1}) (\Theta \, dt - a \, d\phi).
\end{equation}
By integrating the flux of electromagnetic tensor  on the boundary of the spacelike hypersurface, we get the density of electromagnetic charge \cite{Dehghani:2003}
\be
\mathcal{Q}=\frac{q\Theta}{4 \pi G_N}\frac{n(n-1)}{2}
\ee
The time-like Killing vector $\partial_t$ corresponds to the conserved quantity, namely the energy density
\begin{equation}\label{energy density}
\mathcal{E} = \frac{m}{16 \pi G_N} \left[ (n + 1) \Theta^2 - 1 \right],
\end{equation}
and the Killing vector along the azimuthal direction $\partial_\phi$ implies the conservation of angular momentum. The angular momentum density is
\begin{equation}
\mathcal{J} = \frac{n + 1}{16 \pi G_N} \, m \, a \, \Theta.
\end{equation}

\begin{figure}
\centering
\includegraphics[width=0.4\textwidth]{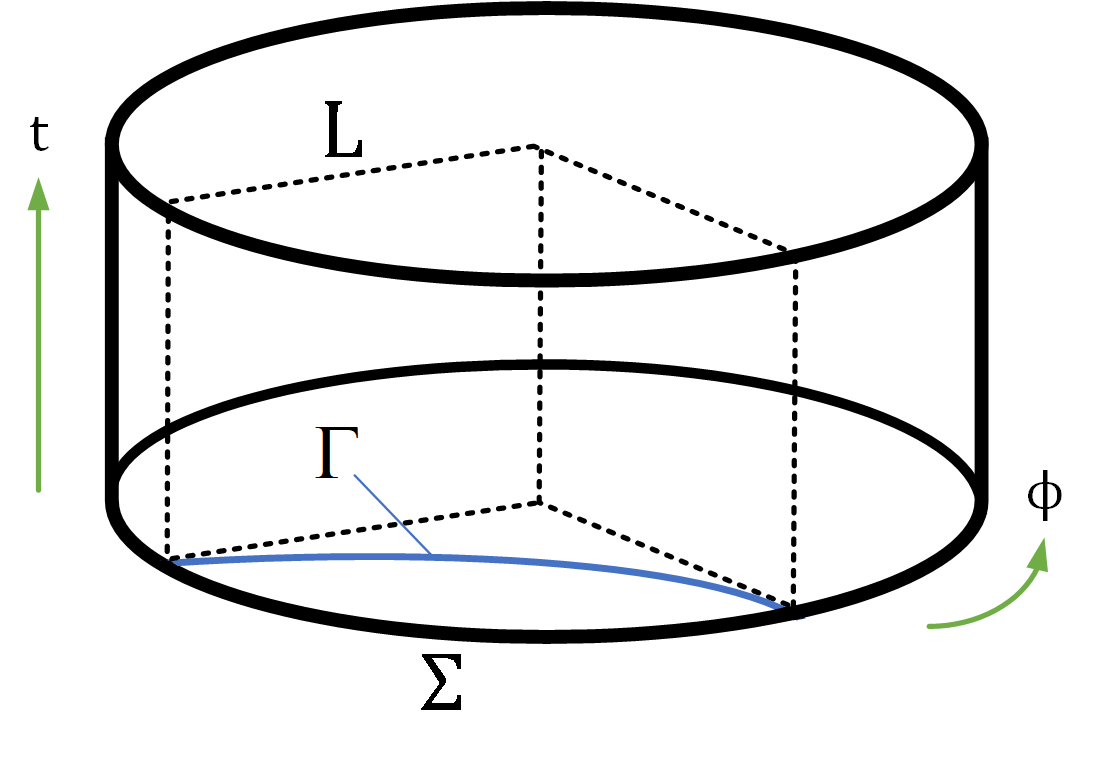}
\includegraphics[width=0.5\textwidth]{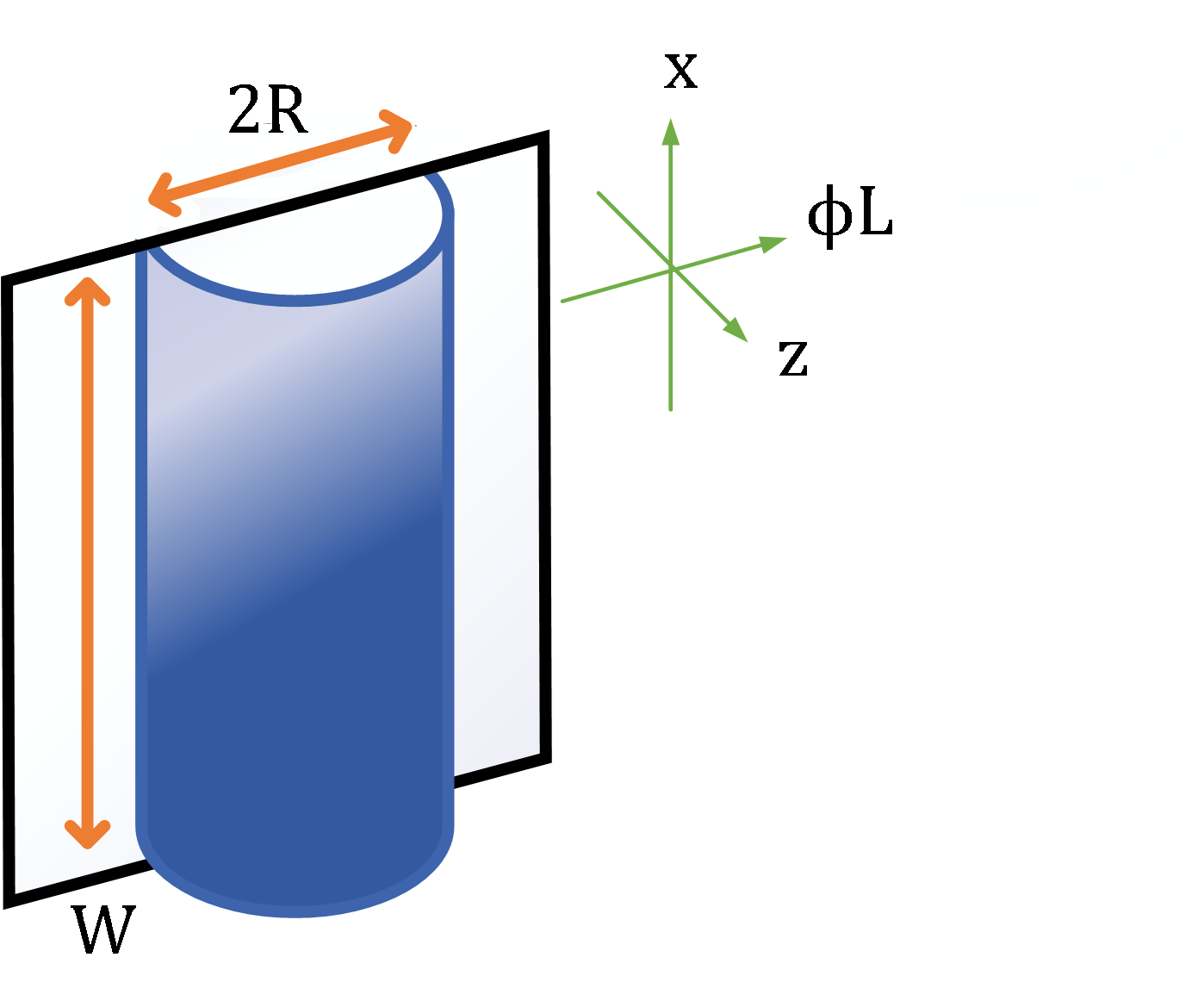}
\caption{{\it Left}: The global structure of $(n+2)-$dimensional cylindrical spacetime. $\Sigma$ is the considered spatial region, $\Gamma_\Sigma$ is the extremal surface (blue curve) for the spatial region $\Sigma$. {\it Right}:  The extremal surface in an $(n+1)-$dimensional spatial slice as $\Sigma$ is a strip.}
\end{figure}

According to the definition of the horizon, $h(z_h) = 0$, we can express mass in terms of horizon radius $z_h$
\begin{equation}
m = \frac{q^2 z_h^{2 n} + 1}{z_h^{n + 1}}.
\end{equation}
For fixing $m$, there are two corresponding horizons, the inner and outer horizons, for the non-extreme case. However, the tip of the extremal surface, $z_b$, cannot greater than the outer event horizon, so we consider $z_h$ as the radius of the outer horizon in the following context. Therefore, we can rewrite the emblackening function~(\ref{hmq}) as
\begin{equation}
h(z) = 1 - g(z), \qquad g(z) = \frac{z^{n+1}}{z_h^{n+1}} + q^2 z^{n+1} \left( z_h^{n-1} - z^{n-1} \right).
\end{equation}
The temperature of the black brane~(\ref{metric}) is
\begin{equation} \label{t}
T = -\frac{h'(z_h)}{4 \pi \Theta} = \frac{n+1}{4 \pi \Theta z_h} \left( 1 - \frac{n-1}{n+1} q^2 z_h^{2 n} \right).
\end{equation}
There is an upper bound of charge parameter
\begin{equation}
q \leq \sqrt{\frac{n + 1}{n - 1}} \frac{1}{z_h^{n}},
\end{equation}
which ensures the non-negative temperature $T \geq 0$. The equality holds when the black brane is extremal. In addition, the electromagnetic field in the bulk geometry corresponds to the chemical potential of the field theory on the boundary
\begin{equation}
\mu = \lim_{z \to 0} A_t(z) = \sqrt{\frac{n}{2 (n - 1)}} \, q \, \Theta \, z_h^{n-1}.
\end{equation}
Thus, we can write the holographic entanglement entropy in terms of chemical potential $\mu$, temperature $T$~\cite{Kundu:2016} and angular velocity $\Omega$.

\section{Integration Forms of Area and Boundary}\label{sec3}
In the following calculations, we will consider a strip boundary defined by
\begin{equation}
\Sigma \equiv \left\{ x_i, \phi \Big| x_i \in \left[ - \frac{W}{2}, \frac{W}{2} \right], \, \phi \in \left[ -\phi_R, \phi_R \equiv \frac{R}{\ell} \right] \right\},
\end{equation}
where $\ell$ is the radius of the cylinder. For simplicity, we let $\ell$ be $1$. The associated extremal surface $\Gamma$ in the bulk can be parameterized by $\xi^i = (\phi, \vec{x})$ with embedding $y^{\mu}(\xi^i) = (z(\phi), \phi, \vec{x})$, and the induced metric is
\begin{equation}
\gamma_{ij} = g_{\mu\nu} \frac{\partial y^{\mu}}{\partial \xi^{i}} \frac{\partial y^{\nu}}{\partial \xi^{j}}.
\end{equation}
Since the bulk metric is independent of the coordinate $\phi$, the extremal surface should have symmetry with respect to $\phi$. Therefore, it is
convenient to let the tip of the extremal surface is at $z_b = z(\phi = 0)$. 

Moreover, one should introduce the UV cutoff $z_\mathrm{UV}$ for regularization of the area of the extremal surface near the boundary $\phi = \phi_R$, i.e. $z(\phi_R) = 0$. Therefore, the area of the extremal surface $\mathcal{A}_{\Gamma}$ is
\begin{eqnarray}
\mathcal{A}_{\Gamma} = L^n\mathcal{A}_{\partial \Sigma} \int_0^{\phi_R - \phi(z_\mathrm{UV})} d\phi \, \mathcal{L}(\dot{z}, z; \phi),
\end{eqnarray}
where $\mathcal{A}_{\partial \Sigma} = 2 W^{n-1}, \, \dot{z} = \partial z/\partial \phi$ and the corresponding Lagrangian is
\begin{equation}
\mathcal{L}(\dot{z}, z; \phi) = \frac1{z^n} \, \sqrt{\frac{\dot{z}^2}{h(z)} + \mathcal{C}(z)}, \qquad \mathcal{C}(z) \equiv \Theta^2 - a^2 h(z) = 1 + a^2 g(z).
\end{equation}
Consequently, the associated Hamiltonian is
\begin{equation} \label{h}
\mathcal{H} = \frac{\partial \mathcal{L}}{\partial \dot{z}} \dot{z} - \mathcal{L} = - \frac{\mathcal{C}(z)}{z^{2n} \mathcal{L}}.
\end{equation}
Since the Hamiltonian does not explicitly depend on $\phi$ implying that the $\mathcal{H}$ is a constant of motion along $\phi$ direction, in the other words, there is a translational symmetry along $\phi$ direction. Because of translational symmetry, it enables us to compute the value of Hamiltonian by evaluating at tip of minimal surface $z = z_{b}$ such that $\dot z = 0$
\begin{equation} \label{hs}
\mathcal{H}(z = z_{b}) = - \frac{\sqrt{\mathcal{C}(z_b)}}{z_b^n} < 0.
\end{equation}
From~(\ref{h}), we can obtain the equation of motion (EoM) for $z(\phi)$ to minimize the area $\mathcal{A}_{\Gamma}$
\begin{equation} \label{zdot}
\dot{z} = \mp \frac{\mathcal{C}(z) \sqrt{h(z)} \sqrt{1 - \frac{z^{2n} \mathcal{C}(z_b)}{z_b^{2n} \mathcal{C}(z)}}}{(-\mathcal{H}) z^n}, \qquad \forall \phi \gtrless 0.
\end{equation}
The equation~(\ref{zdot}) shows that the shape of the minimal surface is symmetric. Therefore, we are able to rewrite the area formula
\begin{eqnarray} \label{area int}
\mathcal{A}_{\Gamma} = L^n\mathcal{A}_{\partial \Sigma} \int_{z_\mathrm{UV}}^{z_b} \frac{1}{z^n \sqrt{h(z)} \sqrt{1 - \frac{z^{2n} \mathcal{C}(z_b)}{z_b^{2n} \mathcal{C}(z)}}} \, dz.
\end{eqnarray}
Also, we can obtain the relation between $\phi_R$ and $z_b$ through
\begin{equation} \label{raphi}
\phi_R= \int_0^{z_b} \frac{(-\mathcal{H}) z^n}{\mathcal{C}(z) \sqrt{h(z)} \sqrt{1 - \frac{z^{2n} \mathcal{C}(z_b)}{z_b^{2n} \mathcal{C}(z)}}} \, dz.
\end{equation}
It is difficult to compute the integration~(\ref{area int}) explicitly. We are going to focus on two specific limits of small and large values of $R$.

\section{\texorpdfstring{Entanglement Entropy in Small $R$ Limit}{Entanglement Entropy in Small R Limit}}\label{sec4}
To evaluate the area of the extremal surface in the small or large $R$ limits, it is convenient to introduce new variables as
\begin{equation} \label{change}
\eta = \frac{z_b}{z_h}, \qquad u = \frac{z}{z_b},
\end{equation}
and then
\begin{equation} \label{bu}
\mathcal{C}(u) = 1 + a^2 g(u), \qquad g(u) = \eta^{n+1} u^{n+1} \left[ 1 + q^2 z_h^{2n} (1 - \eta^{n - 1} u^{n - 1}) \right].
\end{equation}
In the small $R$ limit, the tip of the extremal surface $z_b$ is far away from the position of the horizon $z_h$, that is, $0 < z_b \ll z_h$ or $0 < \eta \ll 1$. Thus we are going to find the leading behavior of area~(\ref{area int}) and $\phi_R$~(\ref{raphi}) in the limit $\eta \rightarrow 0$, respectively.

\subsection{\texorpdfstring{Relation Between $\phi_R$ and $\eta$}{The Relation Between phi R and beta}}
Expanding the integrand of~(\ref{raphi}) in the small $\eta$, we get ($n \ge 2$)
\begin{equation} \label{phi small r beta exp}
\frac{\phi_R}{(-\mathcal{H})} =\int_0^1 \left[ C_{n+1}(u) \eta^{n+1} +\left(C_{2n+2}^{(1)}(u) + C_{2n+2}^{(2)}(u)\right) \eta^{2n+2}+ C_{3n+1}(u) \eta^{3n+1}  + O\left( \eta^{3n+3} \right) \right] du,
\end{equation}
where the first two leading orders are
\begin{eqnarray}
&& C_{n+1}(u) = \frac{u^n z_h^{n+1}}{\sqrt{1 - u^{2n}}},
\\
&& C_{2n+2}^{(1)}(u) = \frac{1 - 2 a^2}2 z_h^{n+1} (q^2 z_h^{2n} + 1) \frac{u^{2n+1}}{\sqrt{1 - u^{2n}}}, \label{c60}
\\
&& C_{2n+2}^{(2)}(u) = \frac{a^2}2 z_h^{n+1} (q^2 z_h^{2n} + 1) \frac{u^{3n} - u^{4n+1}}{(1 - u^{2n})^{3/2}}, \label{c61}\\
&& C_{3n+1}(u) =\frac{\left(a^2-1\right) q^2 u^{3 n} z_h^{3 n+1}}{2 \sqrt{1-u^{2 n}}}.\label{c62}
\end{eqnarray}
We employed the Euler integral of the first kind \footnote{The integration form of beta function is 
$$ \int_0^1 x^{\mu - 1} (1 - x^\lambda)^{\nu - 1} \, dx = \frac{B(\mu/\lambda, \nu)}{\lambda}, \quad \forall \; \mu > 0, \; \nu > 0, \; \lambda > 0, $$
and $B\left( \frac{\mu}{\lambda}, \nu \right) = \frac{\Gamma(\mu/\lambda) \Gamma(\nu)}{\Gamma(\mu/\lambda + \nu)}$,} to evaluate the integration
\begin{eqnarray}
\int_0^1 C_{n+1}(u) \, du &=&\tilde R_0 z_h^{n+1}, \label{c3r}
\\
\int_0^1 C_{2n+2}^{(1)}(u) \, du &=& \left(1 - 2 a^2\right) \left( q^2 z_h^{2n} + 1 \right) \tilde{R}_1 z_h^{n+1} , \label{c61r}
\\
\int_0^1 C_{2n+2}^{(2)}(u) \, du &=& \frac{a^2(n+1)}{2n} (q^2 z_h^{2n} + 1) \left( 4\tilde{R}_1-\tilde{R}_0 \right) z_h^{n+1}. \label{c622r}
\\
\int_0^1 C_{3n+1}(u) \, du &=&\frac{1}{2} \left(a^2-1\right) \left(2-\frac{1}{n+1}\right) q^2 \tilde{R}_0 z_h^{3 n+1}. \label{c63r}
\end{eqnarray}
where $\tilde{R}_0 \equiv \frac{\sqrt{\pi} \, \Gamma\left( \frac12 + \frac1{2 n} \right)}{\Gamma\left( \frac{1}{2 n} \right)}$ and $\tilde{R}_1 \equiv \frac{\sqrt{\pi} \, \Gamma\left( \frac1{n} \right)}{2 n (n + 2)\Gamma\left( \frac12 + \frac1{n} \right)}$. Note that the divergences of both two ``beta-functions''-like contributions in (\ref{c622r}) indeed cancel out to give a finite result.
To compute~(\ref{c61}), we need to convert $(1 - u^{2n})^{-3/2}$ to summation form
\begin{eqnarray}
\int_0^1 \frac{u^{3 n} - u^{4 n + 1}}{(1 - u^{2 n})^{3/2}} du &=& \sum_{j=0}^{\infty} \frac{2 \Gamma\left( j + \frac{3}{2} \right)}{\sqrt{\pi} \, \Gamma(j + 1)} \int_0^1 u^{2 n j + 3 n} (1 - u^{n + 1}) \, du
\nonumber\\
&=& \frac{n+1}{n}\left(4\tilde{R}_1 - \tilde{R}_0\right).
\end{eqnarray}
On top of that, the Hamiltonian in the small $\eta$ limit looks like
\begin{equation} \label{h small}
\mathcal{H} = - \frac{1}{z_h^n \eta^n} - \frac{a^2}{2 z_h^n} (q^2 z_h^{2n} + 1) \eta + \frac{1}{2} a^2 q^2 z_h^n \eta^n + O(\eta^{n+2}).
\end{equation}
Plugging the results~(\ref{c3r}), (\ref{c61r}), (\ref{c622r}), (\ref{c63r}) and~(\ref{h small}) into~(\ref{phi small r beta exp}), we found the $\phi_R$ in the small $\eta$ limit
\begin{align} \label{small phi}
\phi_R &= \tilde{R}_0 z_h \eta + \left[\tilde{R}_1+ \frac{2a^2}{n} \left(\tilde{R}_1-\frac{\tilde{R}_0}4\right) \right] (q^2 z_h^{2n} + 1) z_h \eta^{n + 2} \\
&-\left\{\frac{1}{2} a^2 q^2 \tilde{R}_0 z_h^{2 n+1}+\tilde{R}_1 z_h \left[\frac{a^2}{2 n}\left(\frac{(n+1) \tilde{R}_0}{\tilde{R}_1}-4\right)-1\right] \left(q^2 z_h^{2 n}+1\right)\right\} \eta^{2n+1}+O(\eta^{2n+3})\nonumber. 
\end{align}
In Schwarzschild case~\cite{Saha:2019ado}, $\phi_R=z_h\sum_{j=0}^{\infty}\tilde{R}_j\eta^{j(n+1)+1}$ with $\tilde{R}_j=\frac{1}{2n}\frac{\Gamma\left(j+\frac{1}2\right)\Gamma\left(\frac{1}2+\frac{1}{2n}+\frac{j(n+1)}{2n}\right)}{\Gamma\left(j+1\right)\Gamma\left(1+\frac{1}{2n}+\frac{j(n+1)}{2n}\right)}$. It is easy to see that the correction in (\ref{small phi}) due to rotation $a$ and charge parameter $q$.

\subsection{Area}
Now we move on to the integration of the area. First, we expand the integrand at $\eta \to 0$ limit
\begin{equation} \label{area small r beta exp}
\mathcal{A}_{\Gamma} = L^n\mathcal{A}_{\partial \Sigma} \int_0^1 \left[ \tilde{C}_{-n+1}(u) \eta^{-n+1} + \left(\tilde{C}_2^{(1)}(u) + \tilde{C}_2^{(2)}(u)\right) \eta^2 +\tilde{C}_{n+1}(u)  \eta^{n+1}+ O(\eta^{n+3}) \right] \, du, 
\end{equation}
where
\begin{eqnarray}
&& \tilde{C}_{-n + 1}(u) = \frac{1}{z_h^{n-1} u^n \sqrt{1 - u^{2n}}}\label{divc1},
\\
&& \tilde{C}_2^{(1)}(u) = \frac{1}{2 z_h^{n-1}} (q^2 z_h^{2n} + 1) \frac{u}{\sqrt{1 - u^{2n}}}, \label{tc60}
\\
&& \tilde{C}_2^{(2)}(u) = \frac{a^2}{2 z_h^{n-1}} (q^2 z_h^{2n} + 1) \frac{u^n - u^{2n+1}}{(1 - u^{2n})^{3/2}}. \label{tc61}
\\
&& \tilde{C}_{n+1}(u) = -\frac{\left(a^2+1\right) q^2 u^n z_h^{n+1}}{2 \sqrt{1-u^{2 n}}}. \label{tc62}
\end{eqnarray}
It is a little different from the integration of $\phi_R$. Here, the contribution of the leading order is divergent due to the divergence of integrant at $u = 0$. By introducing the UV cut-off $u_\mathrm{UV} = z_\mathrm{UV}/z_b$, and the relation 
\begin{equation}
\frac1{n - 1} - \int_0^1 \frac1{u^n} \left( \frac1{\sqrt{1 - u^{2 n}}} - 1 \right) du = \frac{\sqrt{\pi} \, \Gamma\left( \frac12 + \frac1{2 n} \right)}{(n - 1) \Gamma\left( \frac1{2 n} \right)}.
\end{equation}
The divergent term comes from~(\ref{divc1}), and we have the ability to extract it
\begin{eqnarray} \label{tc3r}
z_h^{n-1} \int_0^1 \tilde{C}_{-n + 1}(u) \, du &=& \int_{u_\mathrm{UV}}^1 \frac{du}{u^n} + \int_0^1 \frac{du}{u^n} \left( \frac{1}{\sqrt{1 - u^{2 n}}} - 1 \right)
\nonumber\\
&=& \frac1{n - 1} \left( \frac1{u_\mathrm{UV}} \right)^{n-1} - \frac{\tilde{R}_0}{(n - 1)}.
\end{eqnarray}
There other two integrations are similar to the calculation of $\phi_R$, it is easy to get the answers
\begin{eqnarray}
\int_0^1 \tilde{C}_2^{(1)}(u) \, du &=& \frac{q^2 z_h^{2 n} + 1}{2 z_h^{n - 1}} (n+2)\tilde{R}_1 , \label{tc61r}
\\
\int_0^1 \tilde{C}_2^{(2)}(u) \, du &=& \frac{a^2}{2n z_h^{n-1}} (q^2 z_h^{2n} + 1) \left(2(n+2)\tilde{R}_1-\tilde{R}_0 \right). \label{tc622r}
\\
\int_0^1 \tilde{C}_{n+1}(u) \, du &=& -\frac{1}{2} \left(a^2+1\right)q^2 \tilde{R}_1 \eta ^{n+1} z_h^{n+1}. \label{tc63r}
\end{eqnarray}
Plugging the results~(\ref{tc3r}), (\ref{tc61r}), (\ref{tc622r}) and~(\ref{tc63r}) into~(\ref{area small r beta exp}), we obtain the area of the extremal surface in terms of $\eta$ in the small $R$ limit. Since~(\ref{small phi}) tells us the relation between $\eta$ and $\mathcal{\phi}_R$, we can write down the entanglement entropy in terms of $\mathcal{\phi}_R$
\begin{align} \label{ee in small}
&\delta\mathcal{S}_{\Sigma}\equiv \mathcal{S}_{\Sigma}-\mathcal{S}_{\Sigma(0)}=  \frac{L^n\mathcal{A}_{\partial \Sigma}}{4G_N}m\left(\frac{n}{2}+a^2 \right) \frac{\tilde{R}_1}{\tilde{R}_0^2} \phi_R^2\\
&+\frac{L^n\mathcal{A}_{\partial \Sigma}}{4G_N}\left\{m z_h^{1-n} \left[a^2 \left(\frac{(n+1) \tilde{R}_0}{\tilde{R}_1}-4\right)-2 n\right]+n q^2 \left[a^2 \left(\frac{\tilde{R}_0}{\tilde{R}_1}-1\right)-1\right]\right\} \frac{\tilde{R}_1\phi_R^{n+1}}{2 n \tilde{R}_0^{n+1}}+O(\phi_R^{n+3})\nonumber
\end{align}
where $4G_N\mathcal{S}_{\Sigma(0)}/(L^n\mathcal{A}_{\partial \Sigma})=\frac{1}{(n-1)z_\mathrm{UV}^{n - 1}} - \frac{\tilde{R}_0^{n}}{(n - 1) \phi_R^{n-1}}$ is the area of the extremal surface in vacuum. The result recovers the previous the previous studies in~\cite{Bhattacharya:2012mi,Fischler:2013, Chaturvedi:2016kbk, Karar:2018ecr, Saha:2019ado,Kundu:2016,Nadi:2019bqu}. Because AdS (in vacuum) is maximally symmetric spacetime, therefore the spacetime geometry felt by any observer in inertial frame is the same. Here we see that there is no rotation parameter in $\mathcal{S}_{\Sigma(0)}$, and the rotation effect vanishes in metric tensor (\ref{metric}) when $h\rightarrow 1$. Apart from this, if we write back the radius of cylinder $\ell$, we will find that the leading behavior looks likes $\delta\mathcal{S}_{\Sigma} \propto \ell^2{\phi_R}^2 $. The reason why it should be quadratic of ${\phi_R}^2$ is that, once we recognize the entanglement entropy $\delta\mathcal{S}_{\Sigma}$ should proportional to the area of $\partial\Sigma$ and the energy density $\mathcal{E}$ (remember $\mathcal{E}\propto m$ in (\ref{energy density})), we are able to deduce the $\delta\mathcal{S}_{\Sigma}$ should proportional to the quadratic of ${\phi_R}^2$ according to the dimension analysis. The charge parameter appears in the sub-leader because $2n>n+1$ for $n\geq2$. The same behavior occurs in the entanglement wedge cross section(EWCS) of the charged black hole~\cite{Velni:2020,Sahraei:2020}. Note that~(\ref{ee in small}) is valid for both extremal and non-extremal case.

\section{\texorpdfstring{Entanglement Entropy in Large $R$ Limit}{Entanglement Entropy in Large R Limit}}\label{sec5}
In the large $R$ limit, $\eta \to 1$, the tip of the extremal surface should vary close to the event horizon, namely
\begin{equation}\label{large_e}
z_b = z_h (1 - \epsilon), \qquad \text{for $0 < \epsilon \ll 1$}.
\end{equation}
Moreover, we believe that the leading term of the ``regularized'' extremal surface, i.e. after removing the UV divergent, in the large $R$ limit is almost equal to the area of even horizon, therefore we have
\begin{equation} \label{event}
\mathcal{A}_\Gamma \Big|_\mathrm{reg.} \simeq \frac{\Theta \, \phi_R}{z_h} \frac{L^n\mathcal{A}_{\partial \Sigma}}{z_h^{n - 1}} \quad \Rightarrow \quad \frac{\mathcal{A}_{\Gamma}}{L^n\mathcal{A}_{\partial \Sigma}} - \frac{1}{\tilde{z}_\mathrm{UV}^{n-1}} \simeq \frac{\Theta \, \phi_R}{z_h^n}.
\end{equation}
Our goal is to find the leading behavior of~(\ref{area int}) and~(\ref{raphi}) in the large $R$ expansion by assuming the corresponding integrations are dominated by the region $z \simeq z_b$ in the large $R$ limit. We can execute the integration after expanding the integrand in IR regions. This approximation is valid to find the leading term which can be justified by checking the results in the small $a$ and $q$ limits. In this thesis, we only show how to deal with the integration~(\ref{raphi}) in the large $R$ limit. It is easy to verify~(\ref{event}) by using the same method to calculate the integration~(\ref{area int}).

Observing that the leading term of the emblackening factor $h(z)$ near horizon $z = z_h$ for non-extremal is proportional to the black brane temperature by~(\ref{t}), and it is proportional to the second derivative of $h(z)$ when black brane is extremal, i.e
\begin{equation}\label{h_ir}
    h(z) \simeq p z_h^q (1 -\eta u)^q~~~\text{where}~~~\begin{cases}
    q=1,~~p= 4 \pi \Theta T& \text{at $T\neq  0$}\\
    q=2,~~p= \frac{n(n+1)}{z_h^2}& \text{at $T=0$}
    \end{cases}\,.
\end{equation}
Here we do the coordinate transformation by (\ref{change}). Recognizing that the main contribution of integration is from the IR region, we can find the leading behavior by finding the expansion of integrand at $u= 1$ and then carry out the integration.  More precisely, we have
\begin{equation}\label{b_ir}
\mathcal{C}(u) = \mathcal{C}(1) + O(u-1) = \left[\Theta^2 + O(u-1)\right]+ O(\epsilon), \quad \mathcal{C}(1) = \Theta^2 + O(\epsilon),
\end{equation}
and
\begin{equation}\label{main structure}
1 - \frac{z^{2n} \mathcal{C}(z_b)}{z_b^{2n} \mathcal{C}(z)}=1-u^{2n}\frac{\mathcal{C}(1)}{\mathcal{C}(u)} = \left\{r\left(1-u\right)+O\left[\left(u-1\right)^2\right]\right\}+ O(\epsilon)
\end{equation}
where 
\begin{equation}\label{r_ir}
    r=2 n + \frac{a^2 \partial_uh(u)}{\Theta^2}= 2 n - 4\pi z_h T\frac{a^2}{\Theta}+O(\epsilon)\,.
\end{equation}
Note that we have ignored the order $O(\epsilon)$ in (\ref{main structure}). You can convince yourself by considering the order $O(\epsilon)$ in (\ref{main structure}) and integrating it, and find that the $O(\epsilon)$ in (\ref{main structure}) does not contribute to the leading behavior of $\phi_R$. After doing so, via (\ref{h_ir}), (\ref{b_ir}), (\ref{main structure}) and (\ref{r_ir}), we obtain
\begin{equation}\label{int_r_large_r1}
    \phi_R\simeq \frac{z_b}{\sqrt{prz_h^q}\Theta}\int_0^1\frac{1}{(1-\eta u)^{\frac{q}{2}}\sqrt{1-u}}\,du\,.
\end{equation}
To deal with the integration (\ref{int_r_large_r1}), we applied the binomial identity
\begin{equation}
    \frac{1}{(1-\eta u)^{\frac{q}{2}}}=\sum _{j=0}^{\infty } \binom{-\frac{q}{2}}{j}(-1)^j \eta ^j u^j\,,
\end{equation}
then the integration~(\ref{int_r_large_r1}) becomes Euler's integral
\begin{equation}\label{r_ir_sum}
    \phi_R\simeq \frac{z_h}{\sqrt{prz_h^q}\Theta}\sum _{j=0}^{\infty } \binom{-\frac{q}{2}}{j} \frac{\sqrt{\pi } \Gamma (j+1)}{\Gamma \left(j+\frac{3}{2}\right)}(-1)^j \eta ^j  =\frac{2z_h}{\sqrt{prz_h^q}\Theta}\, _2F_1\left(1,\frac{q}{2};\frac{3}{2};\eta \right)\,.
\end{equation}
The summation in (\ref{r_ir_sum}) is exactly equal to the hypergeometric function.\footnote{The hypergeometric function is defined by $$\,_2F_1(a,b;c;z)\equiv\sum _{j=0}^{\infty } \frac{ \left(a\right)_j \left(b\right)_j}{\left(c\right)_j}\frac{z^j}{j!}$$where $$\left(q\right)_j\equiv\frac{\Gamma(q+j)}{\Gamma(q)}$$.} Hence, for finite temperature, (\ref{r_ir_sum}) becomes
\begin{equation} \label{true phi}
\phi_R \simeq - \frac{z_h \, \ln\epsilon}{\sqrt{8 \pi n \Theta z_h \, T}} \tilde{f}(a, q),
\end{equation}
where
\begin{equation}
\tilde{f}(a, q) \equiv \frac{\sqrt{2 n}}{\sqrt{2 n + a^2(n - 1) (1 + q^2 z_h^{2n})}}\simeq \frac{\sqrt{2 n}}{\sqrt{2 n - 4\pi z_h T\frac{a^2}{\Theta}}}.
\end{equation}
Note that the rotation-charge couple $\tilde{f}(a, q)$ has a property $\tilde{f}(0, q) = 1$.
For zero temperature, we concluded that
\begin{equation} \label{ext}
\phi_R \simeq \frac{\pi z_h}{n \Theta \sqrt{2 (n + 1) \epsilon}}.
\end{equation}
Here we have used (\ref{large_e}) to convert $\eta$ to $\epsilon$. It is easy to check that our results (\ref{true phi}) and (\ref{ext}) are consistent with~~\cite{Fischler:2013,Liu:2013una, Chaturvedi:2016kbk,Kundu:2016, Karar:2018ecr}.

\section{Conclusions and Outlooks}\label{sec6}
We derive the holographic entanglement entropy in the small and large $R$ limits in the Kerr-Newman-AdS black holes by considering the shape of the subsystem is a strip. After regularization (to remove the UV divergent term), we discover that the effect of rotation parameter $a$ and mass parameter $m$ always come in quadratic of $\phi_R$ in the small $R$ limit (\ref{ee in small}), and in the large $R$ limit, $q$ and $m$ absorb into the position of horizon $z_h$, and remember $\Theta$ is related to rotation parameter $a$. In fact, it is due to the length contraction of event horizon in the $\phi$ direction under Lorentz boost, that is, $\phi_R\rightarrow \phi'_R=\Theta \phi_R$. Note that both results, (\ref{ee in small}) and (\ref{event}), exhibit the extensive property of entanglement entropy.

The entanglement entropy is an measurement that quantifies how much entanglement between $\Sigma$ and $\Sigma^c$. Let's think about the subsystem in a pure state, there is no uncertainty. Therefore, we never loss any information and the von Neumann entropy vanishes. On top of that, before the observation, we can regard the entanglement entropy $\mathcal{S}_\Sigma$ as the information loss from $\Sigma^c$ is owned by the observer stand in $\Sigma$, and the observer in $\Sigma$ is not accessible to $\Sigma^c$. It indicates that, in non-vacuum state, the two observers standing in $\Sigma$ but with different angular velocity will have different information loss from $\Sigma^c$. Our results indicate that the equilibrium state is only dependent on the thermodynamic variables of the grand canonical ensemble $(\phi_R,\mathcal{E},\mathcal{Q},\mathcal{J})$.

In the future, it may interesting to consider the time depend background, which is modeled by the in-falling dust collapse to form an Kerr-Newman-AdS black holes. The generalization of time dependent proposal is worked by~\cite{Hubeny:2007xt}, and there are several examples of time dependent case~\cite{Liu:2013iza, Liu:2013qca, Albash:2010mv,Alishahiha:2014, Fonda:2014ula,Sun:2021}.



\begin{acknowledgments}
I would like to thank Professor Chiang-Mei Chen for the valuable discussion. The work was supported in part by the Ministry of Science and Technology, Taiwan.
\end{acknowledgments}


\end{document}